\documentclass[a4paper]{mem}
\usepackage{natbib}
\usepackage{graphicx}
\usepackage[a4paper]{hyperref}
\idline{73}{1}
\begin{document}
   \title{Building up the Galactic Halo: The Sagittarius dSph and the
   	globular cluster M22
\thanks{Based on observations made with the European Southern 
Observatory telescopes, using the Wide Field Imager, as part of the 
observing program 65.L-0463. Also based on data obtained from the 
ESO/ST-ECF Science Archive Facility.}
\thanks{This publication makes use of data products from the 
Two Micron All Sky Survey, which is a joint project of the University of 
Massachusetts and the Infrared Processing and Analysis Center/California 
Institute of Technology, funded by the National Aeronautics and Space 
Administration and the National Science Foundation}}

   \author{L. Monaco \inst{1,2}, 
   F. R. Ferraro  \inst{2},\\ 
           M. Bellazzini \inst{1}
          \and
          E. Pancino \inst{1}\fnmsep
}

   \offprints{L. Monaco}
\mail{via Ranzani,1 40127 Bologna, ITALY}

   \institute{INAF - Osservatorio Astronomico di Bologna, 
 via Ranzani,1 40127 Bologna, ITALY \email{monaco,bellazzini,pancino@bo.astro.it}\\ 
              \and  Dipartimento di Astronomia, Universit\`a di Bologna,
	      via Ranzani,1 40127 Bologna, ITALY \email{ferraro@bo.astro.it}
             }

   \abstract{The Sagittarius dwarf spheroidal (Sgr) and a few paculiar Galactic Globular Clusters (GGC) 
   		can be considered as building blocks of the 
   		Galactic Halo. We present a series of results based on a wide field photometry of the Sgr system
		and the GGC M22.
		In particular for Sgr we present: (1) the detection of the RGB-bump, a feature that together with 
		the shape of the RGB can be used to constrain the allowed range of age and metallicity and
		(2) the detection of a clear blue horizontal branch which probes the existance of a metal poor
		population.
		In the case of M22 we present clear evidence that differential reddening is the primary cause for
		the observed spread of the evolutionary sequences. 
	
   }
   \authorrunning{L. Monaco et al.}
   \titlerunning{Sgr dSph \& M22}
   \maketitle
%

\section{Introduction}


There is now a growing wealth of evidence that the Galactic Halo has been, at least partially,
assembled by the slow accretion of subunits. 
This process could be a promising local counterpart of the hierarchical merging acting on cosmological scale 
\citep{free}.

In the context the study of both accreting events (i.e. Sgr) 
and peculiar globular clusters (i.e.
$\omega$~Cen, M22) can be important. Our group is currently involved in the study of both these types of objects.  
In this contribution we will discuss the cases of Sgr and M22. 

The Sgr dSph \citep{igi}
offers the unique opportunity of studing an ongoing process of accretion.
However due to the intrinsic low surface brightness of this distrupting system, an adequate characterization of
the evolutionary features of the stellar populations in the main body of Sgr is still lacking. 
In the following section we will show how
a large area survey (one square degree) centered on the globular cluster M54, which is thought to be the nucleus of Sgr,
reveals several features never seen before in this galaxy.

The globular cluster M22, besides $\omega$~Cen, is the only other globular cluster suspected to
have a significant metallicity spread \citep{hesser} and eventually it could be another remnant of a past merging event. 
In the last section however, we will show that in M22 an important role is played by differential reddening, thus
leaving not much room for a large metallicity spread.   
%

%
%
%

\section {A detailed view of the evolutionary features of the Sgr dSph}
In the following subsection we will illustrate some results based on the wide field photometry 
presented by \citet{bump}, where we have first detected the RGB-bump.

\subsection{The Red Giant Brach Bump}
\begin{figure*}
   \centering
   \resizebox{\hsize}{!}{\includegraphics[width=17truecm,height=17truecm]{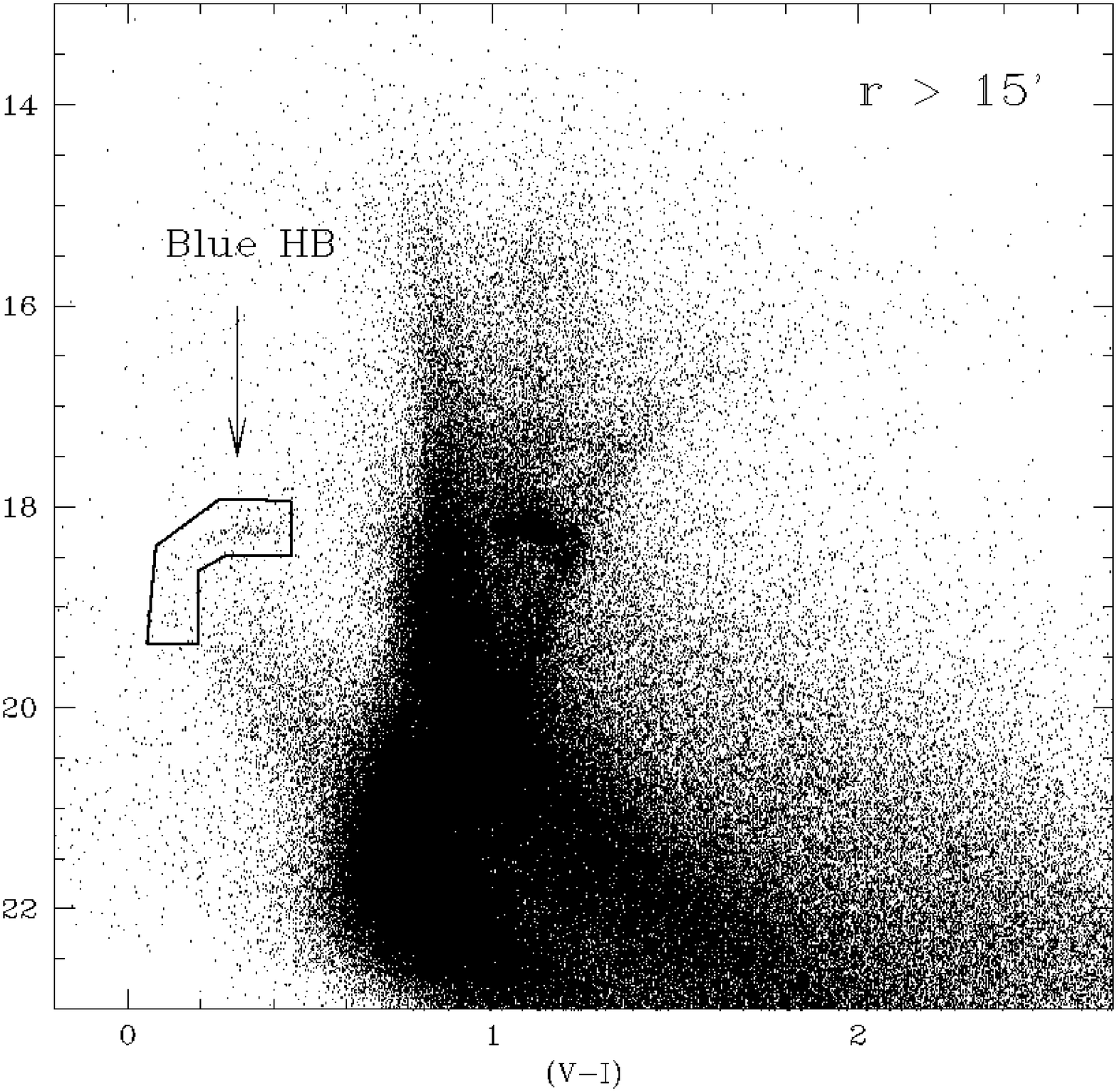}
	{\includegraphics[width=17truecm,height=17truecm]{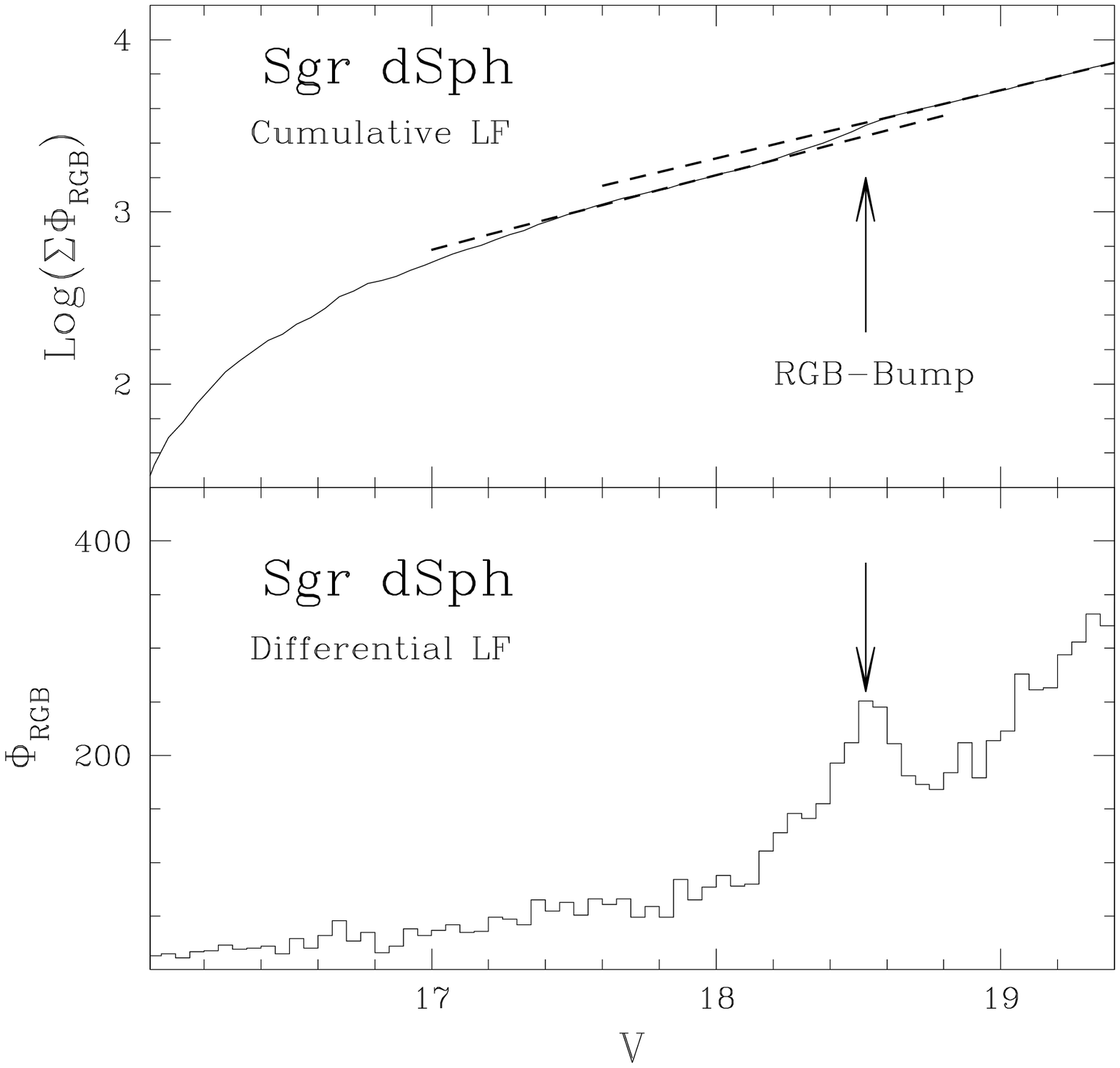}}}     
\caption{Left-hand panel: color-magnitude diagram for the
stars of our samples which are more than 15' far from M~54. Right-hand panels: Differential (lower panel) and cumulative (upper panel) luminosity 
functions for the stars selected in the region of the RGB of the Sgr. 
The arrows indicate the location of the RGB-bump at V=18.55$\pm$0.05. }  
        \label{f2}
    \end{figure*}

We analysed a sample of images obtained at the 
2.2m ESO/MPI telescope at la Silla, Chile, using the {\it Wide Field 
Imager} (WFI) covering a wide 
region of $\sim 1^{\circ} \times 1^{\circ}$ around M~54.
In Figure~\ref{f2} the $(V,V-I)$ CMD for a subsample of the $\sim$490,000 stars measured in the global field 
of view is shown. 
The RGB of the Sgr's metal-rich population (Pop-A, which is the dominant population) 
is easily recognizeable at the red side of 
the diagram, extending to a very red color, $(V-I)\sim 2.7$, with the 
corresponding, well populated red HB clump at $V\sim18.22$. The RGB 
bump of the Sgr can also easily be seen as a clump of stars along the RGB at 
$V\sim 18.5$ and $(V-I)\sim 1.2$.   

In Figure~\ref{f2} (right-hand panels), we show the differential (lower panel) and 
cumulative (upper panel) luminosity functions for the RGB of the Pop-A Sgr (selected from the global sample).
%
The RGB-bump is clearly identified in both panels of figure~\ref{f2}: 
it is located at V$^{bump}$=18.55$\pm$0.05.This is the first 
detection of such a feature in the Sgr dSph.

In the simplest scenario, 
the presence of a single-peaked bump in the differential LF suggest the 
existence of a dominant stellar population, relatively homogeneous 
in metallicity and age, in the inner regions of the Sgr galaxy.

The Pop-A RGB is well fitted by the RGB ridge-line of the 
globular cluster 47~Tuc, that has $[M/H]\simeq -0.6$. However, a 
significant difference in the RGB-bump luminosity between Sgr and 
47~Tuc has been measured (of about 0.12 mag). This fact suggest that Pop A is several Gyr 
younger than 47~Tuc, in good agreement with previous results based on 
the comparison of the Main Sequence Turn Off's \citep{b99b,ls00}. 
On the other hand, since at a fixed metallicity a younger age implies 
bluer RGB colors, we have to conclude that the mean metallicity of 
Pop~A should be higher than that of 47~Tuc \citep[see, e.g.][]{co01}.
 
A full self-consistency among the spectroscopic and photometric 
constraints (including the RGB-bump) is achieved if a mean metallicity 
of $-0.6< [M/H]\le -0.4$ and a mean age of $7\ge age\ge 4$ Gyr are 
assumed for Pop-A.

\subsection{A Blue Horizontal Branch}
All the stars with r$>$15' from the M54 center are plotted in Figure~\ref{f2} (left-hand panel).
A clear blue horizontal sequence can be recognized
at V$\simeq$18.2 and 0.15$\leq$(V-I)$\leq$0.4. Note that these stars cannot be associated to M~54 
\citep[R$_{t}\simeq$7.5 arcmin][]{trager}. Hence the mere detection of this BHB is a strong and
unanbiguous demonstration of the presence of a metal poor population in Sgr \citep[see also][]{ls00,b99b}.

A BHB was never seen in the main body of Sgr before. \citet{new} detected a blue HB in a halo structure which they claim 
to be associated with Sgr. The detection of an extended BHB in the main body of the Sgr 
provides a further confimation that the structure detected by \citet{new} can be associated with the Sgr. 
%

\section {M22: really a peculiar cluster?}
\begin{figure*}
   \centering
   \resizebox{\hsize}{!}{\rotatebox[]{-0}{\includegraphics[clip=true]{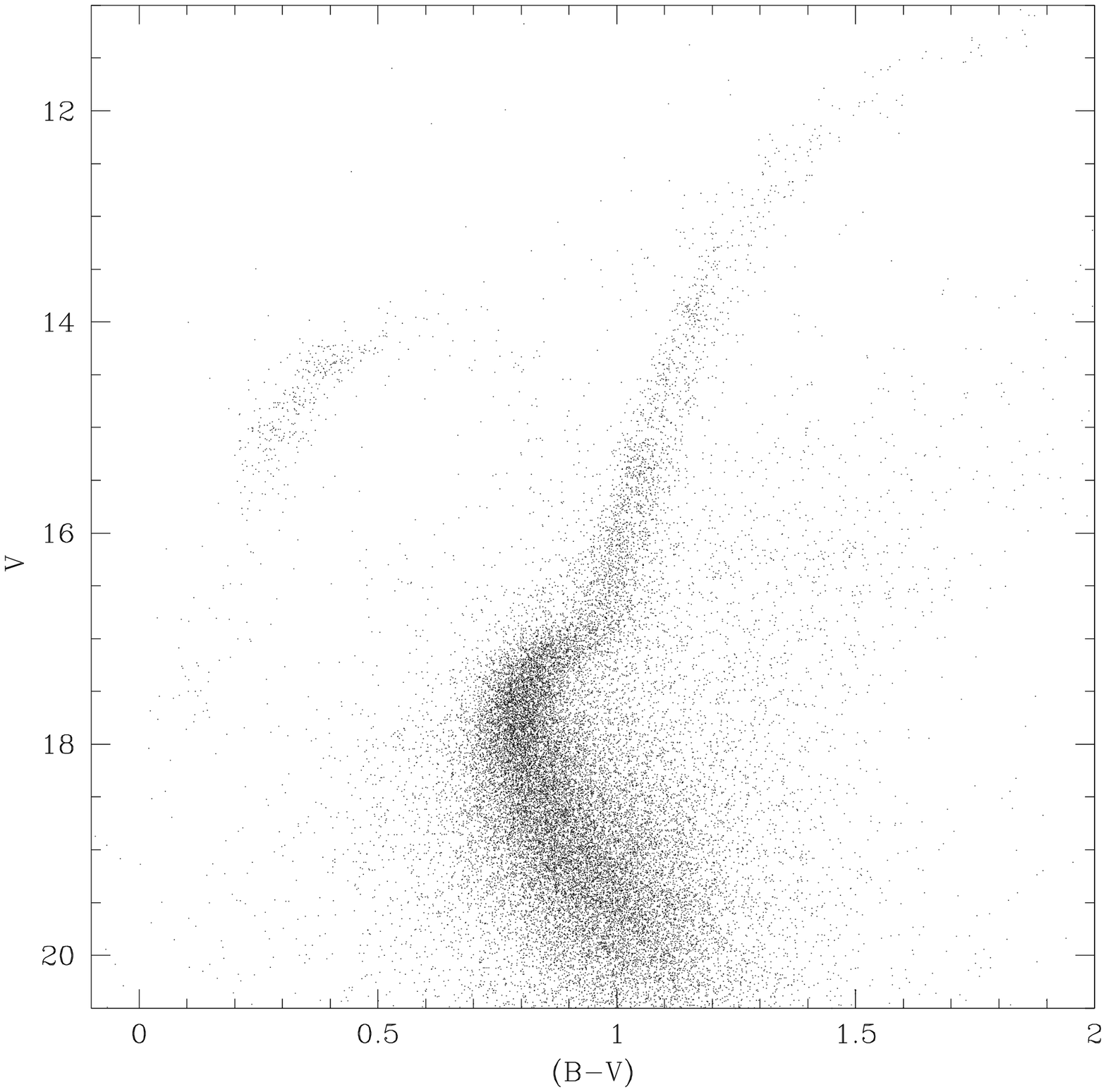}}
	{\includegraphics[clip=true]{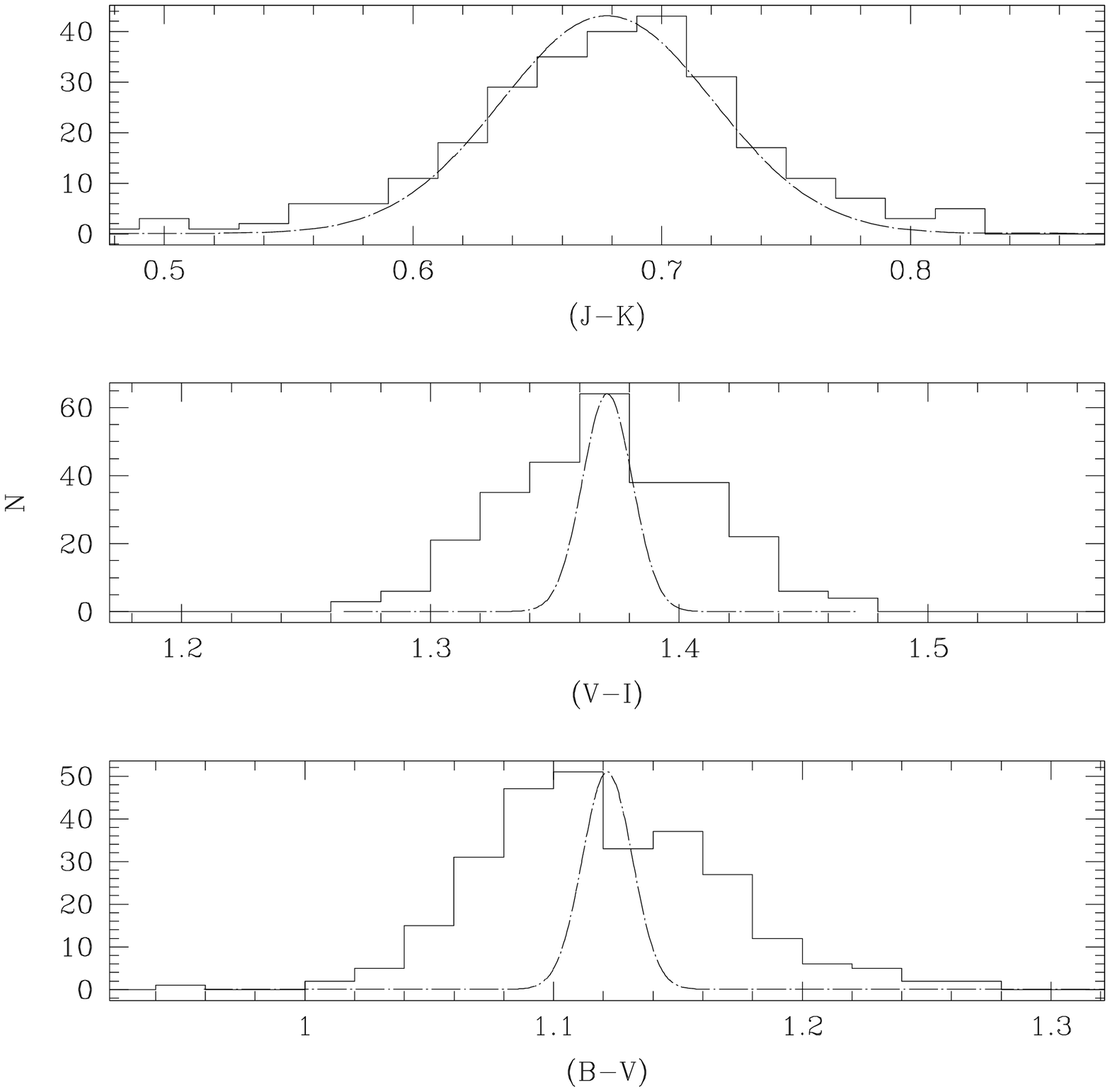}}}	
	\caption{Left panels: color magnitude diagram of M22. Right panels: the width of the RGB is shown in the 
	(B-V), (V-I) and (J-K) colors.}
	\label{dr1}
    \end{figure*}

We again analysed images taken using the WFI, the total area covered is 30'$\times$30'.
The most striking feature which can be seen in the color magnitude diagram of M22 is the large
color spread of the RGB (see figure \ref{dr1}. This color spread could be due either to an internal 
metallicity spread or to the presence of differential reddening, or both.

The presence of differential reddening was already put into evidence in the past (see also 
\citet{rich}).
%
However the presence of differential reddening does not exclude the presence of some (small) degree of
metallicity spread. In order to estimate the amount of differential reddening present in the
region of M22 in figure \ref{dr1} (right hand panels) we compared the RGB color dispersions 
in the (B-V), (V-I) and (J-K) colors. 
%
%
%

On each panel a gaussian curve representing the instrumental
error is also plotted. 
While in the (B-V) and (V-I) color the gaussian is quite different from the real distribution, in the (J-K) color, 
which is the less sensitive to reddening, the gaussian curve approches quite
closely the observed color distribution. 

From these diagrams we evaluate the amount of differential reddening, $\Delta$E(B-V)=0.06$\pm$0.01, and that 
the width of the RGB scales from the optical planes to the infrared plane exactly as expected from 
the reddening law.
This leads us to conclude that an eventual residual metallicity spread must be small.

Until now, no firm conclusion from the spectroscopic studies to the problem of the 
metallicity spread has been drawn for M22. However \citet{leh} found  
a $\Delta[Fe/H]\simeq$0.4 dex for a sample of 10 stars (the larger sample studied so far). 

In order to solve the debate about the metallicity spread of
M~22 we are performing high resolution spectroscopy: 10 spectra of 
RGB stars have been already acquired using UVES@VLT.
Moreover M~22 is one of the target to be observed by the new ESO-VLT multifiber
spectrograph FLAMES using the observing time granted to the Ital-FLAMES 
consortium.
%
%

\begin{acknowledgements}
      Lorenzo Monaco is supported by the Italian Ministero dell'Universit\`a e della ricerca scientifica through
      the grant COFIN-2001028897 assigned to the project {\it The origin and the evolution of Stellar Population
      in the Galactic Spheroid}.
\end{acknowledgements}

\bibliographystyle{aa}

\end{document}